\documentclass[article]{aa}
\usepackage[utf8]{inputenc}
\usepackage[T1]{fontenc}

\usepackage{ulem}
\usepackage{enumitem}
\usepackage{amsmath}
\usepackage{comment}

\usepackage{graphicx}
\usepackage{multirow} 
\usepackage{amssymb}
\usepackage{color}
\usepackage{url}
\graphicspath{{../figures/}}
\usepackage{setspace}

\newcommand\al{\alpha}

\newcommand\gm{\gamma}

\usepackage{natbib,twoopt}
\usepackage[breaklinks=true]{hyperref} 
\bibpunct{(}{)}{;}{a}{}{,}             
\makeatletter
  \newcommandtwoopt{\citeads}[3][][]{\href{http://adsabs.harvard.edu/abs/#3}%
    {\def\hyper@linkstart##1##2{}%
     \let\hyper@linkend\@empty\citealp[#1][#2]{#3}}}
  \newcommandtwoopt{\citepads}[3][][]{\href{http://adsabs.harvard.edu/abs/#3}%
    {\def\hyper@linkstart##1##2{}%
     \let\hyper@linkend\@empty\citep[#1][#2]{#3}}}
  \newcommandtwoopt{\citetads}[3][][]{\href{http://adsabs.harvard.edu/abs/#3}%
    {\def\hyper@linkstart##1##2{}%
     \let\hyper@linkend\@empty\citet[#1][#2]{#3}}}
  \newcommandtwoopt{\citeyearads}[3][][]%
    {\href{http://adsabs.harvard.edu/abs/#3}
    {\def\hyper@linkstart##1##2{}%
     \let\hyper@linkend\@empty\citeyear[#1][#2]{#3}}}
\makeatother

\renewcommand{\linenumbers}{}
\title{Gamma rays from star clusters and implications for the origin of Galactic cosmic rays}
\titlerunning{Gamma and cosmic rays from star clusters}
\authorrunning{P. Blasi}

\author{Pasquale Blasi\inst{1,2}}
\institute{
GSSI - Gran Sasso Science Institute, Viale F. Crispi 7 - I-67100 L’ Aquila, Italy
\and
INFN-Laboratori Nazionali del Gran Sasso, Via G. Acitelli 22, Assergi (AQ), Italy\\
\email{pasquale.blasi@gssi.it}
}

\abstract
{Star clusters are often invoked as contributors to the flux of Galactic cosmic rays and as sources potentially able to accelerate particles to $\sim$PeV energies. The gamma radiation with $E\gtrsim$ TeV recently observed from selected star clusters has profound implications for the origin of Galactic cosmic rays.}
{We show that if the gamma rays observed from the Cygnus cocoon and Westerlund 1 are of hadronic origin, then the cosmic rays escaping the cluster at energies $\gtrsim 10$ TeV must cross a grammage inside the cluster that exceeds the Galactic grammage. At lower energies, depending on the model adopted to describe the production of gamma rays, such grammage may exceed or be comparable with the grammage inferred from propagation on Galactic scales.} 
{The flux of gamma rays is analytically computed for a few models of injection of cosmic rays in star clusters, and compared with the flux measured from selected clusters.}
{In all models considered here, comparing the inferred and observed gamma ray fluxes at $E\gtrsim$ TeV, we conclude that CRs must traverse a large grammage inside or around the cluster before escaping. Clearly these implications would not apply to a scenario in which gamma rays are produced due to radiative losses of leptons in the cluster. Leptonic models typically require weaker magnetic fields, which in turn result in maximum energies of accelerated particles much below $\sim$ PeV.}
{We conclude that if gamma ray emission in SCs is a generic phenomenon and if this radiation is due to hadronic interactions, either star clusters cannot contribute but a small fraction of the total cosmic ray flux at the Earth, or their contribution to the grammage cannot be neglected and the paradigm of Galactic transport should be profoundly revisited.}

\begin{document}
%
\maketitle
\thispagestyle{empty}
\section{Introduction}

Star clusters (SCs) have long been proposed as possible sources contributing to the Galactic cosmic ray (CR) flux. The interest in such sources is mainly due to the fact that SCs might host the right conditions for the acceleration of CRs up to the knee and provide an explanation of the well known $^{22}$Ne/$^{20}$Ne excess in CRs \citep{Gupta+2020}.

Particle acceleration could occur at the termination shock of the collective wind launched by the young stars located in the cluster core, if the SC is sufficiently compact \cite[]{Morlino+2021,BlasiMorlino2023}, or occur due to supernova explosions \cite[]{Vieu+2022}, for sufficiently old SCs. Second order Fermi acceleration in the bubble has also been invoked as a possible mechanism of particle acceleration \cite[]{Vink2024}.
The maximum energy that can be achieved at the termination shock depends rather crucially on details of CR transport in the bubble: for instance, as discussed by \cite{Morlino+2021,BlasiMorlino2023}, although a formal estimate of the maximum energy, $E_{\rm max}$, attainable at the termination shock can easily approach $\sim$PeV, the spherical symmetry of the problem introduces a curvature in the spectrum of accelerated particles that starts at energies much lower than $E_{\rm max}$. The only exception to this conclusion is the case of Bohm diffusion, which however is unlikely to apply to the case of extrinsic turbulence. 

From the observational point of view, gamma ray telescopes such as HAWC \cite[]{2021NatureHAWC} and LHAASO \cite[]{LHAASO-2021Nature} have measured the flux of gamma rays from selected SCs and the spectrum appears to drop appreciably at energies much lower than PeV, despite the fact that photons with energy as high as 2.5 PeV have recently been detected by LHAASO from the a region of $\sim 6^o$ in the direction of the Cygnus OB association \cite[]{LHAASO2024}. Gamma ray emission has now been detected from a number of SCs, such as Westerlund 1 \citep{Abramowski_Wd1:2012}, Westerlund 2 \citep{Yang+2018}, the Cygnus cocoon \citep{Ackermann:2011p3159,Aharonian+2019NatAs}, NGC 3603 \citep{Saha_NGC3603:2020}, BDS2003 \citep{HAWC:2021}, W40 \citep{Sun_W40:2020} and 30 Doradus in the LMC \citep{HESS-30Dor:2015}. Several very young SCs have also been detected in gamma rays in the $\lesssim 100$ GeV energy range \cite[]{Peron2024}.

Both the detailed calculations of particle acceleration and transport in the bubble excavated by the wind, as carried out by \cite{BlasiMorlino2023} (see also \cite{MenchiariThesis2023}), and the more phenomenological estimates presented in \cite{LHAASO-2021Nature}, suggest that in order to account for the level of gamma ray emission in the Cygnus SC, a mean density of $\sim 5-10 \rm cm^{-3}$ is needed in the bubble. Such gas is not the diffuse gas associated with the collective wind of the SC, that has a much lower density, typically $\sim 10^{-3} \rm cm^{-3}$. The mean density should rather be dominated by atomic and molecular gas clumped in dense regions/clouds with larger local density. Such clumps can be either the leftover of the passage of the shock associated with the collective wind, or be formed as a result of the cooling of the gas plowed away toward the outer edge of the bubble. Accepting the presence of this gas, and attributing the gamma ray emission to hadronic interactions with such gas, \cite{BlasiMorlino2024} showed that the spectra of H and He nuclei escaping the SC are different, in agreement with what is inferred from CR data collected by the CREAM collaboration \cite[]{2010CREAM}, by PAMELA \cite[]{PAMELA-P-He}, AMS-02 \cite[]{ams-P,AMS-He}, DAMPE \cite[]{DAMPE-P,DAMPE-He} and CALET \cite[]{CALET-P,CALET-He}. The disappointing by-product of this finding is that in this case heavier nuclei are severely depleted due to spallation inside the bubble, so that SCs cannot contribute an appreciable flux of nuclei heavier than He. 

This was the first symptom of a more general implication of the detection of gamma rays from Cygnus and other clusters: if of hadronic origin, this radiation forces us to conclude that CRs in SCs need to cross a large grammage inside the bubble. For the reference values of the mean density quoted by \cite{LHAASO-2021Nature}, such a grammage exceeds the total one that CRs are expected to accumulate in the Galaxy while diffusing out. 
One could argue that this conclusion may differ by changing the model of CR injection. Hence, below we consider three models for the origin of the gamma ray emission in a star cluster: in the first model, CRs are all accelerated in the center of the SC and diffuse outward while interacting with dense clouds (this is the reference model considered by \cite{LHAASO-2021Nature} for the Cygnus region). In the second model, particle acceleration occurs at the termination shock and particles advect and diffuse outward \cite[]{Morlino+2021}. In the third model, CRs are assumed to be injected in an impulsive event in the center and diffuse outward. The event might be a supernova explosion in the core of the cluster. 

We show that in all models listed above, the local grammage traversed by CRs with $E\gtrsim 10$ TeV while escaping the SC exceeds the one inferred from secondary/primary ratios \cite[]{Evoli1,Evoli2,Schroer2021Nuclei}. The implications for lower energy CRs are also discussed. In the light of this finding, we are forced to conclude that either SCs only contribute a negligible fraction of the CRs we observe at the Earth, or CRs originate in SCs and also accumulate their grammage there rather than in the ISM on Galactic scales. Thi second scenario would require a major revision of the foundations of the model of CR transport in the Galaxy. In alternative, the gamma ray emission that we measure might be due to leptonic processes, which makes the predictions independent of the gas density. These models (see for instance \cite{Harer2023} for one such models applied to Westerlund 1) require lower strength of the magnetic field in the bubble, which in turn implies that SCs cannot reasonably play the role of PeVatrons. Finally, it might be that the SCs that have been detected so far in gamma rays are unusual and their characteristics are irrelevant for the bulk of the SCs. Although this cannot be excluded, it would definitely be surprising.

If the constraints to the grammage that are inferred here are meaningful for the bulk of SCs, then one should keep them in mind even when SNRs are considered as the main source of CRs, since the majority of such sources are expected to be located in SCs.  
From the observational point of view, an important clue to the origin of the gamma radiation and the constraints on grammage might come from the detection of gamma rays from very young SCs, for which we can be reasonably sure of the absence of SN events taking place at the present time.

The article is organised as follows: in Sect. \ref{sec:models} we discuss the gamma ray emission in the three models of CR production illustrated above. In Sect. \ref{sec:grammage} we elaborate on the implications of the gamma ray measurements for CR grammage, comparing the results with standard models of CR transport in the Galaxy. In Sect. \ref{sec:discuss} we outline our conclusions.

\section{Models of gamma ray production in SCs}
\label{sec:models}
As we discuss below, the effect of the grammage is so macroscopic that a sophisticated treatment of pp interactions is not necessary. Moreover, in the analysis below, we will focus on gamma ray energies $\gtrsim 1$ TeV, where scaling relations are sufficiently accurate for our purposes here. We assume that gamma rays with given energy $E_\gamma$ are produced by protons with energy $E$ such that $E_\gamma=\eta E$, with $\eta\simeq 0.1$. The cross section for inelastic $pp$ interactions is considered as constant with energy, $\sigma_{pp}=32$ mb.

Here we consider three models of CR transport in the SC. In Model 1 we assume, following \cite[]{LHAASO2024}, that a source in the center of the SC continuously injects CRs with a spectrum $Q(E)=A (E/m_p)^{-\alpha}$, with $\alpha\gtrsim 2$. We also follow \cite[]{LHAASO2024} in assuming that transport is purely diffusive with diffusion coefficient $D(E)$. In the energy region of interest here, $E_\gamma\gtrsim 1$ TeV, the diffusion equation can be used in its stationary form, so that the CR density of energy $E$ at distance $r$ from the center is 
\begin{equation}
    n_{CR}(r,E)\approx\frac{Q(E)}{4\pi r D(E)}.
\end{equation}
The injection spectrum is normalized to a fraction $\xi_{CR}$ of the luminosity $L_{SC}$ of the SC (the one that involves plasma motion and can be seed of particle acceleration), through
\begin{equation}
    Q(E)=\xi_{CR}\frac{L_{SC}(\al-2)}{(m_p c^2)^2} \left( \frac{E}{m_p}\right)^{-\al},
\end{equation}
where we assumed that the energy in the form of CRs is dominated by relativistic particles, namely $E\gtrsim m_p$. Within a factor of order unity this is a good approximation for $2\lesssim\al\lesssim 3$, as expected from diffusive shock acceleration (DSA). These source spectra are also compatible with those found in \cite[]{LHAASO2024} to provide a good fit to the observed gamma ray emission for the Cygnus region. Notice that, quantitatively, including in the normalization procedure also particles with energy lower than $m_p$, would strengthen the conclusions drawn below. 

In the second model (Model 2), we follow \cite{Morlino+2021,BlasiMorlino2023} in assuming that particle acceleration takes place at the termination shock of the collective wind. The structure of the bubble is properly described by \cite{Morlino+2021}: the location of the termination shock is at $r=R_{sh}$, while the bubble extends to $r=R_b$, but as we will see below we do not need to write down these quantities explicitly. Particle acceleration is assumed to transform a fraction $\xi_{CR}$ of the luminosity of the wind, $L_w=(1/2) \dot M v_w^2$, into CRs. Here $\dot M$ is the rate of mass loss in the form of a collective wind and $v_w$ is the wind velocity. The structure of the bubble is such that the shocked wind, namely the plasma behind the termination shock has a velocity $v(r)=(v_w/4)(r/R_{sh})^{-2}$, assuming that the termination shock is strong (compression factor of 4). The density downstream of the shock is constant. In the wind region (upstream of the termination shock), the velocity is constant to the value $v_w$, while the density is 
\begin{equation}
    \rho_w(r)=\frac{\dot M}{4\pi r^2 v_w}.
\end{equation}
Assuming again that the density of accelerated particles at the termination shock has the shape of a power law, $n_{CR}(E)=A (E/m_p)^{-\alpha}$, and that the energy in the form of accelerated particles carries a fraction $\xi_{CR}$ of $L_w$, it is easy to derive:
\begin{equation}
    n_{CR}(E)=2\xi_{CR}\frac{L_w (\al-2)}{4\pi R_{sh}^2 (m_p c^2)^2}\left(\frac{E}{m_p}\right)^{-\al}.
\end{equation}
In the following we will assume that this density remains constant downstream of the termination shock: while this is a better approximation at $E\lesssim 1$ TeV than at higher energies, where escape from the bubble becomes dominated by diffusion rather than advection, it also maximises the gamma ray emission, making the results illustrated below even stronger. 


In the third model (Model 3) the particles are assumed to be produced in the core of the SC, as in Model 1, but the source is impulsive. This situation might mimic the case of a SN explosion. It is not clear how likely it is to have had a SN explosion in Cygnus or Westerlund 1, although in the Cygnus region there is evidence for a pulsar in the core, with an estimated age of $\sim 10^5$ yrs. Since there is no evidence of a shell in the bubble, we assume that the SN event occurred indeed $\gtrsim 10^5$ years ago. Once produced, CRs are assumed to diffuse outwards and eventually escape. 

If $E_{SN}$ is the kinetic energy of the SN ejecta, then the density of particles in the bubble can be estimated as
\begin{equation}
    n_{CR}(E)\approx \frac{\xi_{CR}E_{SN}(\alpha-2)}{(m_p c^2)^2 [4\pi D(E) t]^{3/2}} \left( \frac{E}{m_p}\right)^{-\al},~~\rm r<\sqrt{4 D t}. 
    \label{eq:nimpulse}
\end{equation}
Notice that the condition that particles of energy $E$ are still inside the SC at time $t$ imposes an upper limit on the diffusion coefficient $D(E)$:
\begin{equation}
    D(E)<7.7\times 10^{27} \rm cm^2/s \left( \frac{t}{10^5 \rm yr} \right)^{-1}\left( \frac{R_b}{100 \rm pc} \right)^2.
    \label{eq:LowD}
\end{equation}
Notice also that the particles at time $t$ fill uniformly a sphere of radius $\sqrt{4 D t}$, while their density drops $\propto \exp[-r^2/4 D t]$ at larger distances. For simplicity we assume that the particles are all concentrated with the diffusion distance. 

\subsection{Gamma ray emission in Model 1: diffusion model}
\label{sec:Model1}
In the following we will denote as $n_{gas}$ the mean gas density in the bubble, to be interpreted as the volume averaged gas density, dominated by the clumpy structure of atomic and molecular gas (see discussion in \cite[]{BlasiMorlino2024}). We will assume that this density is constant throughout the bubble, since the available information on the spatial distribution of the gas is very poor (see also discussion in \cite[]{MenchiariThesis2023}). 

As discussed above, in Model 1 \cite[]{LHAASO2024}, we assume that a source in the SC core injects CRs at a constant rate and these particles move outward only due to diffusion. In this simple picture there is no bulk motion of the plasma, which mimics a situation in which the SC is unable to launch a collective wind (although even in that case it is difficult to imagine that a collective motion may be completely absent). As argued by \cite{Vieu2024}, this might be the case of the Cygnus cocoon.

The emissivity in the form of gamma rays can be written as:
\begin{equation}
    J_{\gm} (E_\gm,r) dE_\gm = n_{CR}(E,r) dE n_{gas} c \sigma_{pp}.
\end{equation}
Using the assumption that $E_\gm = \eta E$, one can write
\begin{equation}
    J_{\gm} (E_\gm,r) E_\gm^2 = \xi_{CR}\frac{L_{SC}(\al-2)}{4\pi r D(E)} \eta^{\al-1}\left( \frac{E_\gm}{m_p}\right)^{-\al+2} n_{gas} c \sigma_{pp}.
\end{equation}
It follows that the gamma ray flux from the entire bubble can be estimated as
$$
    E_\gm^2 \Phi_\gm (E_\gm) =\int_0^{R_b} dr \frac{r^2}{d^2} \xi_{CR}\frac{L_{SC}(\al-2)}{4\pi r D(E)} \eta^{\al-1}\left( \frac{E_\gm}{m_p}\right)^{-\al+2}  
$$
\begin{equation}
    \times n_{gas} c \sigma_{pp} = \xi_{CR}\frac{L_{SC}(\al-2)}{4\pi d^2 D(E)} \eta^{\al-1}\left( \frac{E_\gm}{m_p}\right)^{-\al+2} \frac{R_b^2}{2} n_{gas} c \sigma_{pp}.
    \label{eq:gammaM1}
\end{equation}
The escape time from the bubble in a purely diffusive case is $\tau_{b}=R_b^2/6D(E)$, so that we can introduce the grammage traversed by particles while escaping the bubble as
\begin{equation}
    X(E)=n_{gas} m_p c \tau_b(E) = n_{gas} m_p c \frac{R_b^2}{6D(E)}.
\end{equation}
This allows us to rewrite Eq. \ref{eq:gammaM1} as:
\begin{equation}
    E_\gm^2 \Phi_\gm (E_\gm) = \frac{3 \xi_{CR} L_{SC}(\al-2)\eta^{\al-1}}{4\pi d^2} \left( \frac{E_\gm}{m_p}\right)^{-\al+2} \frac{X(E)}{X_{cr}},
    \label{eq:Model1_gamma_X}
\end{equation}
where $X_{cr}=m_p/\sigma_{pp}$ is the critical grammage for $pp$ interactions. As written here, this expression can be also used, within factors of order unity, when the gas is not uniformly distributed, for instance when the whole gas is concentrated in a thin shell at the edge of the bubble. 

\subsection{Gamma ray emission in Model 2: acceleration at the termination shock}\label{sec:Model2}

With the same assumptions on pp collisions adopted above, the gamma ray emissivity in the case of particle acceleration at the termination shock can be written as:
\begin{equation}
    J_{\gm} (E_\gm) = 2\xi_{CR}\frac{L_w (\al-2)}{4\pi R_{sh}^2 (m_p c^2)^2} \eta^{\al-1}\left( \frac{E_\gm}{m_p}\right)^{-\al} n_{gas} c \sigma_{pp}.
\end{equation}
The integral over the downstream of the termination shock, the region that dominates the gamma ray emission in this model, leads to:
\begin{equation}
    E_\gm^2 \Phi_\gm (E_\gm)  = 2\frac{\xi_{CR} L_w (\al-2)}{4\pi R_{sh}^2} \eta^{\al-1}\left( \frac{E_\gm}{m_p}\right)^{-\al+2} \frac{R_b^3}{3 d^2} n_{gas} c \sigma_{pp}.
    \label{eq:gammaM2}
\end{equation}
In order to maximize the gamma ray flux predicted by the model and make our conclusions more solid, we assume here that particle escape from the bubble is dominated by advection. As discussed by \cite{BlasiMorlino2024}, at energies $\gtrsim 1$ TeV escape becomes dominated by diffusion, but this leads to a smaller effective emission region for gamma rays and correspondingly smaller gamma rays ray flux, hence our assumption is conservative in terms of deriving constraints on the grammage inside the bubble. Recalling that the plasma velocity drops downstream of the termination shock, we can write:
\begin{equation}
    \frac{dr}{dt} = \frac{v_w}{4}\left( \frac{r}{R_{sh}} \right)^{-2} \to \tau_{adv}\approx \frac{4}{3}\frac{R_{sh}}{v_w} \left( \frac{R_b}{R_{sh}} \right)^{3},
\end{equation}
where for simplicity we assumed that $R_b\gg R_{sh}$. Replacing this expression in Eq. \ref{eq:gammaM2}, we easily obtain:
\begin{equation}
    E_\gm^2 \Phi_\gm (E_\gm) \approx \frac{2 \xi_{CR} L_w (\al-2)\eta^{\al-1}}{4\pi d^2} \left( \frac{E_\gm}{m_p}\right)^{-\al+2} \frac{X(E)}{X_{cr}},
    \label{eq:Model2_gamma_X}   
\end{equation}
where the grammage is defined as $X(E)=n_{gas}m_p c \tau_{adv}$. A quick comparison between Eqs. \ref{eq:Model1_gamma_X} and \ref{eq:Model2_gamma_X} immediately suggests that Models 1 and 2 are going to lead to very similar constraints in terms of grammage. 

\subsection{Gamma ray emission in Model 3: Impulsive injection of CRs}\label{sec:Model3}

The detection of gamma rays of energy $E_\gamma$ implies that the parent CR particles, with energy $E=E_\gamma/\eta$ are still confined inside the SC. This trivially implies that the gamma ray spectrum integrated over the volume is insensitive to the diffusion coefficient. Hence the flux of gamma rays can be written using Eq. \ref{eq:nimpulse} integrating on $r\leq \sqrt{4 D(E) t}$:
\begin{equation}
    E_\gm^2 \Phi_\gm (E_\gm) = \frac{\xi_{CR} E_{SN}(\al-2)\eta^{\al-1}}{3 \pi^{3/2} d^2} \left( \frac{E_\gm}{m_p}\right)^{-\al+2} n_{gas} \sigma_{pp} c,
    \label{eq:Model3}
\end{equation}
with the usual meaning of the symbols. It is worth noting how this result, contrary to the ones obtained for the other two models, depends only upon the density of target gas and not on the grammage. On the other hand, as discussed above (see Eq. \ref{eq:LowD}), the condition that particles are still trapped in the bubble after a time $t$ will allow us to infer a lower limit to the grammage traversed by particles before they are released into the ISM. 

\section{Constraints on the grammage inside the Cygnus cocoon and Westerlund 1}
\label{sec:grammage}

The LHAASO experiment has recently measured the gamma ray flux with energy above $\sim 1$ TeV from the Cygnus cocoon at the level of $E_\gm^2 \Phi_\gm (E_\gm)\simeq 6\times 10^{-11}\rm erg\,cm^{-2}\,s^{-1}$, over a region that extends $\sim 150$ pc from the center of Cygnus SC. The emission is claimed to have hadronic origin, and it is difficult to imagine otherwise given the extended morphology. The slope of the spectrum is $\sim 2.7$ (at $E\gtrsim 1$ TeV) and probably reflects the energy dependent diffusion coefficient, together with the shape of the spectrum near the maximum energy. This point is however of little interest here, in that we focus on the lowest energy bin, at 1 TeV, and investigate the dependence of our constraints on the value of $\al$, the slope of the injection spectrum. The HESS telescope has observed gamma rays from the direction of the SC Westerlund 1 \cite[]{Aharonian2022}, with $E_\gm^2 \Phi_\gm (E_\gm)\simeq 1.6\times 10^{-11}\rm erg\,cm^{-2}\,s^{-1}$ at $E_\gamma=1$ TeV. In the following we will adopt a distance of 1.4 kpc for Cygnus and 3.9 kpc for Westerlund 1.

In Model 1, we normalize the luminosity to $L_{SC}=10^{39}L_{39}$erg/s. Notice that photons with energy $E_\gm$ are produced by protons with energy $E=10$ TeV, hence we will be able to constrain the grammage traversed in the SC by protons with 10 TeV energy. The gamma ray flux computed using Eq. \ref{eq:Model1_gamma_X} can be written as $E_\gm^2 \Phi_\gm\approx 1.5\times 10^{-9} \xi_{CR}L_{39}d_{kpc}^{-2} X(E)\rm\,erg\,cm^{-2}\,s^{-1}$ ($4.8\times 10^{-10} \xi_{CR}L_{39}d_{kpc}^{-2}X(E)\rm\,erg\,cm^{-2}\,s^{-1}$) for $\alpha=2.2$ ($\alpha=2.4$). Comparing this prediction with the flux as measured by LHAASO for the Cygnus cocoon we get a grammage $X(E=10TeV)\approx 0.08 \xi_{CR}^{-1}L_{39}^{-1}\,\rm g\,cm^{-2}$ ($0.25 \xi_{CR}^{-1}L_{39}^{-1}\,\rm g\,cm^{-2}$) for $\alpha=2.2$ ($\alpha=2.4$).

\begin{figure*}
\centering
\includegraphics[width=.48\textwidth]{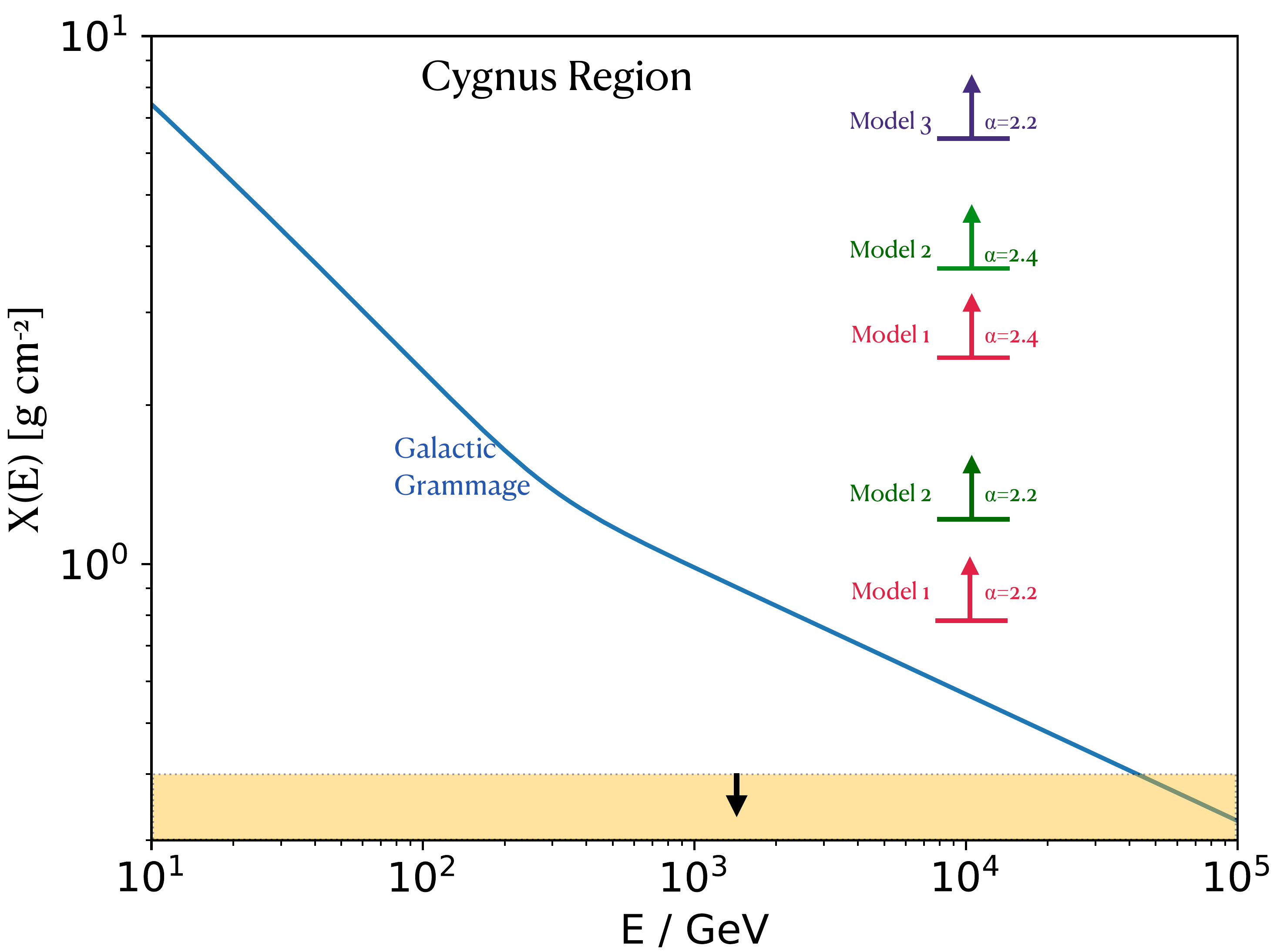}
\includegraphics[width=.48\textwidth]{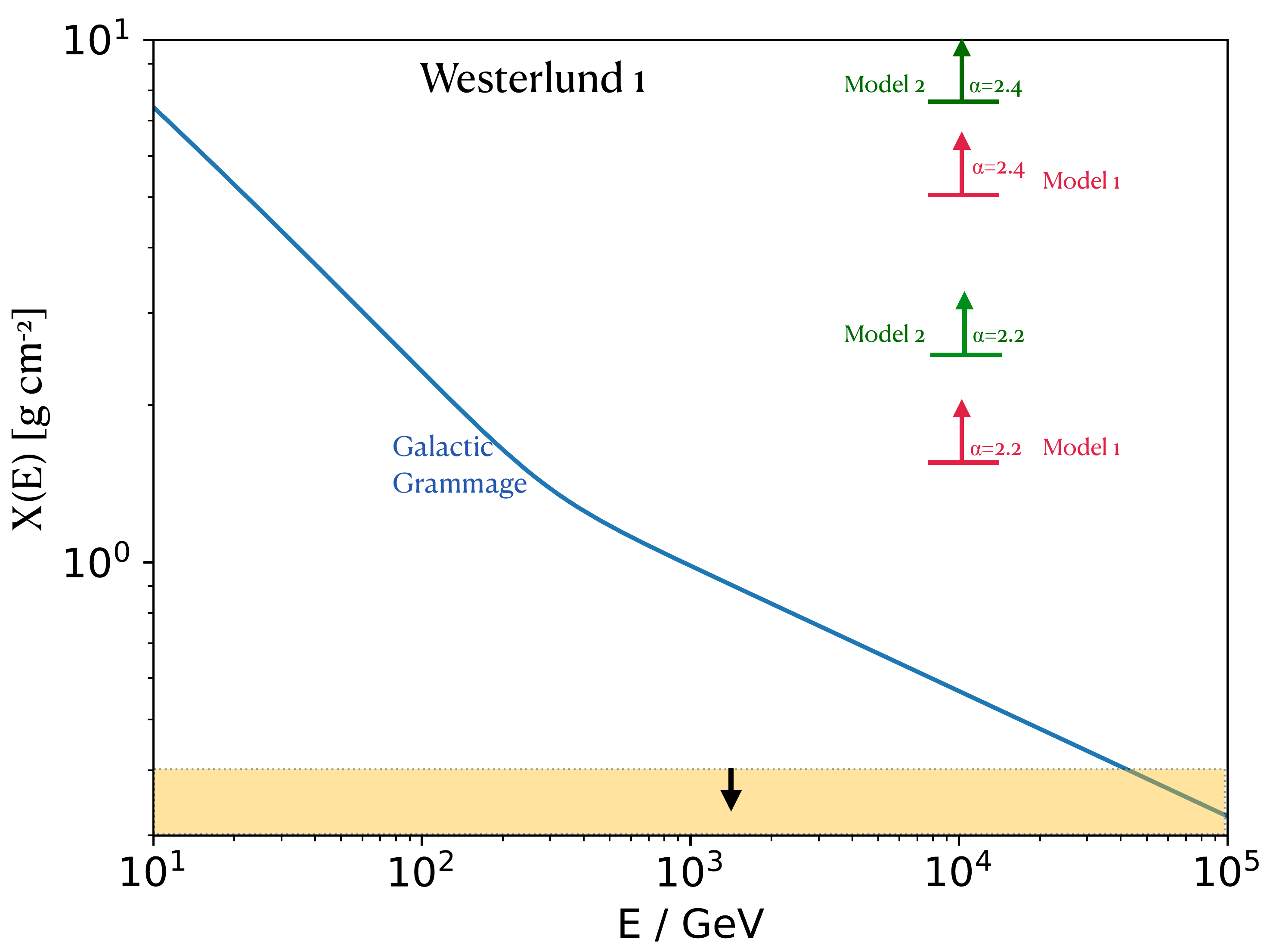}
\caption{Lower limits to the grammage inferred for the Cygnus region (left) and Westerlund 1 (right) if the observed gamma rays are of hadronic origin, in three models of gamma ray production as described in the text. These limits are shown together with an up-to-date computation of the Galactic CR grammage inferred from secondary/primary ratios \cite[]{Schroer2021Nuclei}. The yellow shadowed region shows the maximum source grammage still compatible with secondary to primary ratios ($\simeq 0.4\,\rm g\,cm^{-2}$).}
\label{fig:Grammage}
\end{figure*}

Finally, requiring a reasonable efficiency of particle acceleration, say $\xi_{CR}\lesssim 0.1$, we immediately infer that in order to account for the gamma ray flux observed by LHAASO, the grammage required for Model 1 has to be $X(E=10 TeV) \gtrsim 0.8 L_{39}^{-1}\,\rm g\,cm^{-2}$ for $\al=2.2$ and $X(E=10 TeV) \gtrsim 2.5 L_{39}^{-1}\,\rm g\,cm^{-2}$ for $\al=2.4$. A similar procedure leads to a lower limit to the grammage in the case of Westerlund 1. Both sets of limits are plotted in Fig. \ref{fig:Grammage} for Cygnus (left panel) and Westerlund 1 (right panel). The figure also shows the grammage traversed by CRs during their journey through the Galaxy, as inferred from secondary/primary ratios (see for instance \cite[]{Evoli1,Evoli2,Schroer2021Nuclei}). The yellow shaded region shows the largest grammage (the so-called source grammage) that would still be compatible with the measured ratios. Interestingly, this grammage is not too far from the one expected to be accumulated while CRs are downstream of the shock of a supernova remnant \cite[]{Aloisio2015}. Any grammage in excess of $\sim 0.4 \rm g~cm^{-2}$ would force us to revisit the very foundations of the picture of CR transport in the Galaxy.  

In fact, this seems to be the case for both Cygnus and Westerlund 1, if the gamma ray emission is of hadronic origin: Model 1 leads to expect a grammage in both cases appreciably larger than the Galactic grammage at 10 TeV. Even more concerning, in Model 1 this grammage would extend to lower energies as a power law, so that the conclusions above would also apply to lower energies, where the grammage is even better constrained. 

Finally, notice that the limit above have been specialized to a SC luminosity of $10^{39}$ erg/s, which is probably somewhat larger than the actual luminosity of the two SCs considered here. For lower values of the luminosity, the limits would become correspondingly stronger. 

In Model 2 we normalize the wind luminosity to $L_{w}=10^{39}L_{w,39}$erg/s. The gamma ray flux computed using Eq. \ref{eq:Model2_gamma_X} can be written as $E_\gm^2 \Phi_\gm\approx 10^{-9} \xi_{CR}d_{kpc}^{-2} L_{w,39}X(E)\rm\,erg\,cm^{-2}\,s^{-1}$ ($3.2\times 10^{-10} \xi_{CR}d_{kpc}^{-2} L_{w,39}X(E)\rm\,erg\,cm^{-2}\,s^{-1}$) for $\alpha=2.2$ ($\alpha=2.4$). Following the same procedure outlined for Model 1, and requiring efficiency $\xi_{CR}\lesssim 0.1$, we infer the corresponding lower limits on the grammage, as shown in Fig. \ref{fig:Grammage} (green symbols for Model 2). The problem discussed above with the grammage appears to be even more severe for Model 2.


Notice that the limits derived for Model 2 should be considered as absolute lower limits to the grammage, since we neglected the diffusive escape of CRs from the cocoon at $E\gtrsim 1$ TeV \cite[]{BlasiMorlino2024}, that would reduce the high energy CR density and increase the required grammage. 

These limits are in line with the results presented by \cite{BlasiMorlino2023,BlasiMorlino2024}, where a gas density $n_{gas}\sim 5-10\,\rm cm^{-3}$ was used, in line with the values inferred by the \cite{LHAASO2024}. Using the expressions above, this would correspond to a CR acceleration efficiency at the termination shock $\xi_{CR}\sim 0.5\%$, in agreement with the values found by \cite{BlasiMorlino2023}. This however implies a grammage in the cocoon of $\sim 50\,\rm g\, cm^{-2}$, well in excess of the Galactic grammage. Not surprisingly, such a grammage was found to be responsible for the complete spallation of heavy nuclei and a slight difference between the escaping spectra of H and He nuclei \cite[]{BlasiMorlino2024}. SCs with characteristics similar to those of the Cygnus cocoon and Westerlund 1 cannot account for the bulk of Galactic CRs unless CR transport is deeply revised. Alternatively, the gamma ray emission may not be due to hadronic interactions, as suggested by \cite{Harer2023} for Westerlund 1. 

In Model 3, we test the possibility that CRs have been injected in the bubble during an impulsive event, a time $t$ in the past, and that the gamma ray emission is being produced by such particles today. The condition that CRs of energy $E$ are still confined inside the bubble leads to an upper limit to the diffusion coefficient, $D(E)<R_b^2/4t$. The comparison between Eq. \ref{eq:Model3} and the observed gamma ray flux from a SC allows us to impose a lower limit to the gas density in the bubble, if we require $\xi_{CR}\leq 0.1$. For Cygnus this limit reads $n_{gas}>n_{min}=62~\rm cm^{-3}$ for $\alpha=2.2$ and $n_{gas}>n_{min}=131~\rm cm^{-3}$ for $\alpha=2.4$. The combination  of these two conditions implies that the grammage must satisfy: 
\begin{equation}
X(E)>\frac{2}{3} n_{min} m_p c t = 1~\rm g~cm^{-2} \left( \frac{n_{min}}{10 \rm cm^{-3}}\right)\left( \frac{t}{10^5 \rm yrs}\right).  
\end{equation}
One can see that not only the values of density in this model appear extreme and hard to reconcile with the formation of a bubble, but also the inferred grammage exceeds the Galactic grammage, more so than in Models 1 and 2 discussed above. In Fig. \ref{fig:Grammage} we only show the limit on the grammage for Model 3 for the cases in which such limit is lower than $10~\rm g~cm^{-2}$. Similar considerations apply to the case of Westerlund 1. 

\section{Discussion}
\label{sec:discuss}

SCs have long been thought to play an important role in the framework of the origin of Galactic CRs, in at least two ways: first, by providing enough CRs to justify the excess of $^{22}\rm Ne$ observed in the cosmic radiation. Second, by possibly providing the right conditions for acceleration up to the knee, something that SNRs seem to struggle to do. The recent detection of selected SCs in high energy gamma rays has fueled renewed interest in this problem, especially in the perspective of SCs playing the role of PeVatrons. As discussed by \cite{Morlino+2021,BlasiMorlino2023}, it seems unlikely that SCs may efficiently accelerate particles up to the knee energy at the termination shock generated by the collective wind. It cannot yet be excluded that higher energies may be reached in SN explosions occurring inside a SC. This latter scenario would apply to SCs that are old enough to host SN explosions. 

Here, we discussed the implications of the recently observed gamma ray emission from selected SCs in terms of their contribution to the bulk of Galactic CRs. In particular, we have shown that the gamma ray emission observed by LHAASO from the direction of the Cygnus cocoon \cite[]{LHAASO-2021Nature,LHAASO2024} and by HESS from the direction of Westerlund 1 \cite[]{Aharonian2022} lead to an estimate of the grammage accumulated by particles while leaving the bubble region, that exceeds the Galactic grammage at $\sim 10$ TeV. The way that this finding extends to lower energies depends on the model adopted to describe the gamma ray emission. Even if this grammage were energy independent, the accumulated grammage at low energies would require substantial revision of the transport of Galactic CRs, in order to accommodate the excess grammage.



If to take for granted the average gas density in the Cygnus cocoon, $n_{gas}=5-10\,\rm cm^{-3}$, as inferred in much of the recent literature \cite[]{Aharonian+2019NatAs,LHAASO-2021Nature,MenchiariThesis2023,Menchiari2024}, the inferred grammage would be $X\sim 25-50\,\rm g\,cm^{-2}$, larger than the total grammage traversed by CRs at $\sim$GeV energies, by about one order of magnitude. Not surprisingly, the calculations of \cite{BlasiMorlino2024}, applied to the Cygnus cocoon, imply that nuclei heavier than He are basically destroyed inside the cocoon, and even for He the spallation reactions are sufficiently severe to harden the He spectrum with respect to H nuclei. 

It is clear that the discovery of high energy gamma ray emission from SCs and most notably from Cygnus and Westerlund 1 represents a milestone in our field of investigation, and a possible clue to the origin of high energy CRs. But the same detection also implies that either SCs cannot contribute but a small fraction of Galactic CRs, or our understanding of how CRs accumulate grammage in the Galaxy needs to be deeply revised. Clearly these conclusions are based on assumptions that may be flawed: for instance, one could speculate that the SCs detected in gamma rays are not representative of the whole class of SCs, implying that perhaps most SCs are not bright in gamma rays. The other possibility is that the gamma radiation observed so far is of leptonic origin, so that it becomes uncorrelated with the grammage traversed in the source. 

Notice that, if most SCs are hadronic gamma ray emitters, as discussed by \cite{MenchiariDiffuse2024}, they would contribute significantly to the diffuse Galactic gamma ray background. On the other hand, while there is a substantial correlation between unidentified Fermi-LAT sources and young SCs, no such strong correlation arises in the TeV range or with SCs that are older than a few million years \cite[]{PeronUnassociated2024}, suggesting that Cygnus and Westerlund might be exceptional. Clearly both these issues require more attention if to assess the role of SCs as potential CR sources.

Models of the origin of Galactic CRs in which grammage is accumulated near sources were actually developed by \cite{Cowsik1975} and recently revised by \cite{CowsikReview2022} as a possible way to address the issue of a rising positron fraction. In general these models struggle to accommodate the features observed in the spectra of primary nuclei and the decay of $^{10}$Be. Nevertheless, it is worth keeping in mind the possibility that at least part of the grammage inferred from measurements at the Earth may be accumulated in or around the sources rather than {\it en route} to the Earth. 

\begin{acknowledgements}
This work has been partially funded by the European Union - Next Generation EU, through PRIN-MUR 2022TJW4EJ and by the European Union - NextGenerationEU under the MUR National Innovation Ecosystem grant  ECS00000041 - VITALITY/ASTRA - CUP D13C21000430001. The author is grateful to E. Amato, C. Evoli, S. Menchiari, G. Morlino, G. Peron and E. Sobacchi for useful conversations.
\end{acknowledgements}

\bibliographystyle{aa}
\bibliography{biblio}
%
\end{document}